\documentclass[aps,pre,twocolumn,preprintnumbers,amsmath,amssymb]{revtex4}
\usepackage{amsmath,amssymb,color,epsfig,latexsym,natbib,graphicx,dcolumn, bm} 

\newcommand\BV{{Brunt-V\"ais\"al\"a frequency}}            
\begin{document}
\title{Vertical drafts and mixing in stratified turbulence: sharp 
       transition with Froude number}

\author{F. Feraco$^{1,2}$, R. Marino$^1$, A. Pumir$^3$, L. Primavera$^2$, P.D. Mininni$^4$, A. Pouquet$^{5,6}$ and D. Rosenberg$^7$ }

\affiliation{
$^1$Laboratoire de M\'ecanique des Fluides et d'Acoustique, CNRS, \'Ecole Centrale de Lyon, Universit\'e Claude Bernard Lyon~1, INSA de Lyon, F-69134 \'Ecully, France.\\
$^2$Dipartimento di Fisica, Universit\`a della Calabria, Italy.\\
$^3$\'Ecole Normale Sup\'erieure de Lyon, Lyon France.\\
$^4$Departamento de F\'{\i}sica, Facultad de Ciencias Exactas y Naturales, Universidad de Buenos Aires, and IFIBA, CONICET, Buenos Aires 1428, Argentina.\\
$^5$Laboratory for Atmospheric and Space Physics, University of Colorado, Boulder, CO 80309, USA.\\
$^6$National Center for Atmospheric Research, P.O.~Box 3000, Boulder, CO 80307, USA.\\
$^7$1401 Bradley Drive, Boulder CO 80305, USA.}

\begin{abstract}
We investigate the large-scale intermittency of vertical velocity and 
temperature, and the mixing properties of stably stratified turbulent flows 
using both Lagrangian and Eulerian fields from direct numerical simulations, 
in a parameter space relevant for the atmosphere and the oceans. Over a range 
of Froude numbers of geophysical interest ($\approx 0.05-0.3$) we observe very 
large fluctuations of the vertical components of the velocity and the 
potential temperature, localized 
in space and time, with a sharp transition leading to non-Gaussian wings of the 
probability distribution functions. This behavior is captured by a simple model 
representing the competition between gravity waves on a fast time-scale and 
nonlinear steepening on a slower time-scale. The existence of a resonant regime 
characterized by enhanced large-scale intermittency, as understood within the 
framework of the proposed model, is then linked to the emergence of structures 
in the velocity and potential temperature fields, localized overturning and 
mixing. Finally, in the same regime we observe a linear scaling of the mixing 
efficiency with the Froude number and an increase of its value of roughly one 
order of magnitude.
\end{abstract}

\maketitle
\section{Introduction}
Intermittency is a hallmark of fully developed turbulence in fluids.
Contrary to the predictions of the Kolmogorov's original theory \cite{K41}, both 
experiments and numerical simulations show that dissipation exhibits intense 
(intermittent) fluctuations, localized in space and time, giving rise 
to a characteristic behavior of quiescence interrupted by local bursts of high 
amplitude \cite{kolmogorov_62,frisch_80}. 
This phenomenon, known as small-scale intermittency, is widely observed in the 
atmosphere \cite{fritts_13b}, where it can explain the formation of rain droplets 
\cite{falkovich_07,Bodenschatz}, and in the ocean in the form of highly concentrated 
and sporadic dissipation \cite{klymak_08,pearson_18}.                                                                                
Small-scale intermittency is often characterized by the 
strong deviations from Gaussian statistics of the probability distribution 
functions (PDF) of velocity and temperature gradients. 
Intermittency, however, is not only present at the smallest scales. In 
transitional pipe flows \cite{barkley_15}, in the problem of mixing of a passive 
scalar by a turbulent flow~\cite{PSS91}, in the solar wind \cite{marino_12}, 
and for stratified flows as in the Earth's atmosphere \cite{chau} and in 
the oceans, non-stationary energetic bursts at scales comparable to that of 
the mean flow are also observed~\cite{mahrt_89,dasaro_11,rorai_14,pearson_18}. 
The origin of this large-scale intermittency in stratified turbulence and the 
mechanisms by which it may affect overturning and mixing in geophysical 
flows are still unclear. In the present letter we characterize the vertical 
drafts occurring at a large scale, using direct numerical simulations (DNS) 
of stably stratified Boussinesq flows, varying the buoyancy frequency. 
{The large scale intermittent behavior is studied here by direct integration 
of the Boussinesq equations, without any parametrization of the smaller scales 
and with periodic boundaries, contrary to what was done in previous work}.    
In particular, we relate the wings of the PDFs of the vertical component of 
Lagrangian and Eulerian velocities and of the potential temperature,
 to large-scale bursts, which result  from the interplay of gravity waves and 
turbulent motions in a range of Froude numbers relevant to geophysical flows. 
This is explained with the help of a simple one-dimensional (1D) model, which 
captures the extreme events \cite{rorai_14} and the sharp transition observed 
in our simulations between Gaussian and non-Gaussian PDFs. 
Finally we provide clear evidence of the connection between local overturning 
events and mixing in stratified flows $-$ characterized through {the} gradient 
Richardson number \cite{Rosenberg_15}, {the} mixing efficiency and the ratio 
of kinetic to potential energy $-$ {with} the emergence of large-scale 
intermittency and structures. 
 
\section{Equations, parameters, and runs}

The Boussinesq approximation assumes that the variations of density are small 
and depend linearly on temperature; they are neglected, except in the expression 
of the buoyancy force. The incompressible velocity field $\mathbf{u}$ 
($\nabla \cdot \mathbf{u} = 0$) and temperature, expressed in suitable units 
satisfy: 
\begin{eqnarray}\label{bequationV} 
  \partial_t \textbf u+(\textbf u \cdot\nabla)\textbf u  
   &=& - \nabla p - N\theta\hat{z} +  \textbf F +\nu\nabla^2\textbf u \\
   \partial_t\theta+\textbf u\cdot\nabla\theta &=& Nw + \kappa\nabla^2\theta \, 
\label{bequationT} \end{eqnarray}
where 
$\theta$ is a temperature fluctuation relative to the mean $\overline\theta$,
$\nu$ is the kinematic viscosity, and  $\kappa=\nu$ is the thermal diffusivity. 
The velocity field is written as $\textbf u=({\bf u}_\perp,w)$ and the 
flows in this study are subject to a random isotropic mechanical forcing
$\textbf F$ (as also described in \cite{marino_14}), applied in a wavenumber 
shell: $k_F=2\pi/L_f \in [2,3]$, the size of the periodic three-dimensional 
computational box being $L_0=2\pi$. Finally, 
$N=\left[-g\partial_z\overline\theta/\theta_0\right]^{1/2}$ is the \BV.
%
\begin{table*} \centering
\setlength\tabcolsep{5.0pt}
\small
\begin{tabular}{c c c c c c c c c c c c c c c c c c c  p{1cm}}
Id&1&2&3&4&5&6&7&8&9&10&11&12&13&14&15&16&17\\
\hline
\hline
$Re$ $(\times 10^3)$ &  3.8 &   3.8 &  3.8 &  3.8 &  3.8 &  3.8 &  3.9 &  3.8 &  {\bf 3.8} &  3.8 &  3.7 &  3.6 &  3.0 &  2.6 &   2.6 &   2.8 &    2.9 \\
$Fr$           &  .015 &   .026 &  .030 &  .038 &  .044 &  .051 &  .068 &  .072 & {\bf .076} &  .081 &  .098 &   .11 &   .16 &  .19 &  .28 &  .56 &  .93 \\
$K_{\tilde w}$ &  3.11 &   3.17 &  3.07 &  3.06 &  3.17 &  3.41 &  7.40 &  9.50 & {\bf 10.44} &  9.39 &  8.86 &  5.63 &  3.87 & 3.53 & 3.30 & 2.95 & 3.00 \\
$\ell_{Oz}$    &   .07 &   .014 &  .018 &  .024 &  .029 &  .037 &  .056 &  .061 & {\bf .067} &  .076 &   .11 &   .14 &   .39 &  .65 & 1.25 & 3.66 & 8.00 \\
$R_B$   &   .9 &   2.5 &  3.4 &  5.6 &  7.3 & 10.2 & 17.7 & 19.7 & {\bf 22.1} & 25.2 & 35.9 & 47.5 & 75.2 & 90.9 & 201  & 895  & 2560 \\
\hline
\end{tabular}
\caption{{\it Table of the runs. Reynolds, Froude and buoyancy Reynolds numbers 
  are respectively $Re$, $Fr$ and $R_B$; $\ell_{Oz}$ is the Ozmidov 
  scale; $K_{\tilde w}$ is the kurtosis of the Lagrangian PDFs computed over 
  the entire integration interval of the particles. 
  Run 9 (boldface) has the highest $K_{\tilde w}$ and is used as a reference case 
  for the visualizations in Figures \ref{f:RENDER} and \ref{f:RENDER2}}.
}
\label{t:runs} 
\end{table*}
We define the Reynolds and Froude numbers, the dimensionless parameters of the 
problem: $Re = UL / \nu , \  Fr = U / LN$,
where $U,L$ are respectively the characteristic velocity and the integral scale 
of the fluid. The buoyancy Reynolds number,
$R_B \equiv Re \, Fr^2 = \Big( \ell_{Oz}/\eta \Big)^{4/3}$, where $\ell_{Oz}
\equiv(\varepsilon_V/N^3)^{1/2}$ and $\eta\equiv(\varepsilon_V/\nu^3)^{-1/4}$ 
are the Ozmidov and Kolmogorov dissipation lengths, and 
$\varepsilon_V \equiv \nu \langle (\nabla \mathbf{u})^2 \rangle$ 
is the kinetic energy dissipation rate. $R_B$  measures the 
relative strength of buoyancy and dissipation: for $R_B=1$, the Ozmidov 
scale at which dispersive and nonlinear effects balance is at the Kolmogorov 
scale.
%
To solve these equations numerically we use the Geophysical High-Order Suite for 
Turbulence (GHOST) code, a versatile pseudo-spectral framework parallelized 
with a hybrid MPI/OpenMP method \cite{hybrid_11}.
 All computations are performed on isotropic grids of $512^3$ points, altogether 
for of the order of 40 $\tau$, where $\tau=L/U$ is the turn-over time. Together 
with the Eulerian velocity and temperature field, we determined also the 
trajectories of tracer particles in the flow, whose positions 
${\bf X}({\bf X}_0, t)$ satisfy $\partial_t{\bf X}({\bf X}_0, t)={\bf u}({\bf X}({\bf X}_0, t),t)$ 
(with $\mathbf{X}(\mathbf{X}_0, 0) = \mathbf{X}_0$). We followed
 $n\approx 1.5\cdot10^6$ particle trajectories, initially uniformly distributed in 
space. 
They are injected randomly after turbulence becomes fully developed, {\it i.e.},
after the dissipation in the flow has reached its maximum (see Fig.~\ref{f:KUR}), 
and their trajectories are determined for a duration of 6 to 10 turnover times. 
Lagrangian temporal statistics are always collected throughout the entire time
of integration. To characterize the statistical distribution of a 
fluctuating variable, we introduce the dimensionless fourth-order moment 
(kurtosis), defined as:
\begin{equation}
\label{kurtomega}
K_{\alpha}=\langle \alpha^4 \rangle / \langle \alpha^2 \rangle^2 \\  
\end{equation}
where $\alpha$ is the generic field and averages can be taken over the entire 
box, horizontal planes, time, or over ensembles of particles as specified below. 
The Gaussian reference value of the kurtosis is equal to 3. 
 \begin{figure*}
 \begin{center}
 \includegraphics[width=17.3cm]{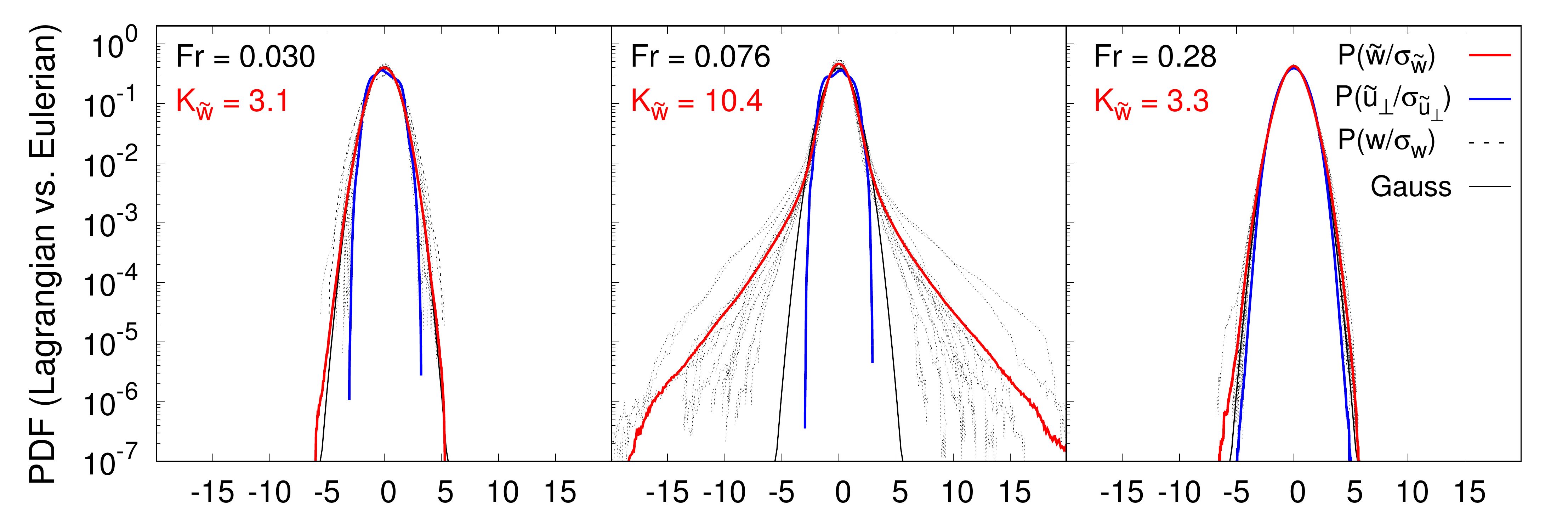}
 \end{center}
\vskip-0.2truein   
 \caption{{\it PDFs of the Lagrangian vertical velocity $\tilde w$ (solid line, 
   red), Lagrangian horizontal
   velocity $\tilde u_{\perp}$ (solid line, blue) and instantaneous Eulerian
   vertical velocity $w$ at several times (all dashed lines, black), for
   runs 3 (left), 9 (middle) and 15 (right). All quantities are normalized by 
   their dispersion. A Gaussian PDF is indicated as reference in each panel
   (solid line, black).}}
  \label{f:PDF}     
\end{figure*}
The Reynolds number varies roughly by $35\%$ throughout the parametric 
study, from $\approx 2600$ to $\approx 3900$. All the runs have been initialized 
with zero potential temperature and a random velocity field with energy 
distributed on spherical shells centered on the wavenumber $k_0=2\pi/L_0$ in the 
range $k_0=[2,3]$. 
%
\section{Large scale intermittency in stratified flows}
As a result of the stratification, the Lagrangian particles mostly wander around 
in horizontal planes, with sharp vertical excursions which leave a signature in 
the PDFs of  $w$ and $\tilde w$, respectively the vertical component of the 
Eulerian velocity field and the vertical velocities of the Lagrangian particles. 
Figure \ref{f:PDF} gives both the instantaneous PDFs of $w$, each corresponding 
to a dotted curve (computed every 1000 time-steps of the DNS, within the domain 
of integration of the particles), and PDFs of the Lagrangian particles velocity 
$\tilde w$  (solid red curves), for three different Froude numbers. In all plots, 
the Lagrangian PDFs appear, up to statistical errors, as the time average of the 
instantaneous Eulerian PDFs. 
In the case shown in the middle panel ($Fr=0.076$), the wings for high values of 
$w$ and $\tilde w$ are significantly broader than the Gaussian wings, with a 
kurtosis of the Lagrangian particles velocity $K_{\tilde w}\approx 10.44$. 
This is the signature of the occurrence of extreme events at some time in the 
evolution of the flow, in some regions of the spatial domain. 
The departures from the Gaussian behavior are very weak for the cases with 
smaller and larger Froude number displayed in the lateral panels ($Fr=0.03$ and 
$Fr=0.28$, respectively left and right), with the kurtosis becoming 
$K_{\tilde w} \approx 3$ for {the} smallest and highest $Fr$ considered in this 
parametric exploration (see Table \ref{t:runs}).
%
 \begin{figure}
 \begin{center}
 \hspace*{-0.5cm}
 \includegraphics[width=9.26cm]{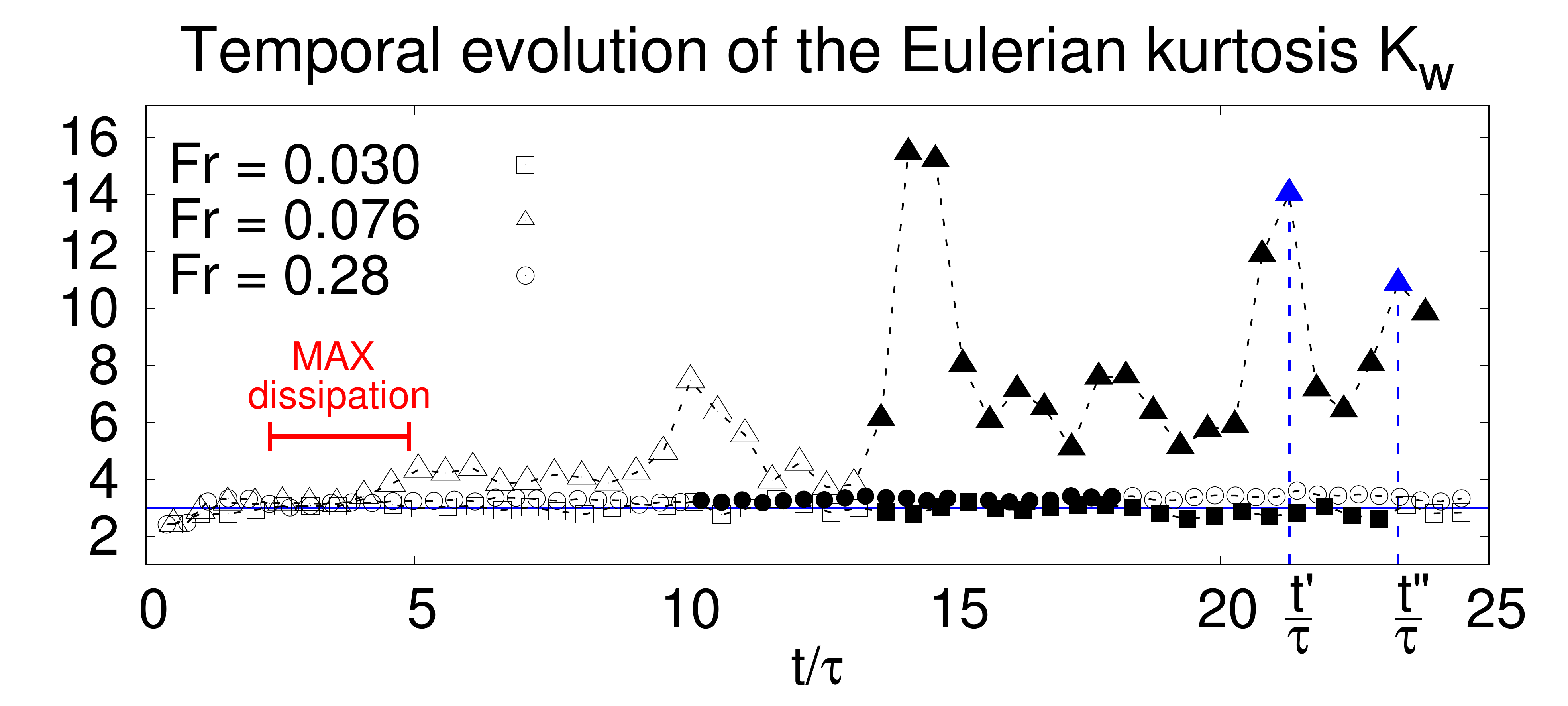}   
 \end{center}
 \vskip-0.2truein   
 \caption{
        {\it Kurtosis of the Eulerian vertical velocity $(w)$ as a function 
        of time for runs 3,9 and 15. Each point is the statistical moment 
        (eq.~\ref{kurtomega}) of the corresponding instantaneous PDFs showed 
        in Fig.\ref{f:PDF} (dashed lines). 
        Solid symbols identify the periods of integration of the Lagrangian 
        particles, injected systematically after the peak of the dissipation, 
        here indicated with a red segment for the runs displayed. Points in 
        blue identify the relative maxima at which the Eulerian fields of 
        run 9 $(Fr = 0.076)$ are rendered in Figures \ref{f:RENDER} and 
        \ref{f:RENDER2}.}  
        }   
 \label{f:KUR}
 \vskip-0.15truein       
 \end{figure}
%
Fig.~\ref{f:PDF} also shows the PDFs of the horizontal Lagrangian components 
of the velocity field $\tilde u_{\perp}$ (solid blue line) that is markedly 
below the Gaussian distribution, in all 17 runs.
For some of the stratified flows under study, the instantaneous PDFs of the
Eulerian vertical velocity $w$ vary strongly over time, as it is for run 9
in Fig.~\ref{f:PDF} (dotted lines, middle panel).  
This time-dependence leads to strong fluctuations of the corresponding Eulerian 
PDFs' kurtosis, $K_w$, shown in Fig.~\ref{f:KUR} ($Fr=0.076$). 
%
%
The kurtosis of the Lagrangian PDFs, $K_{\tilde w}$, resulting from global
spatio-temporal statistics of $1.5\cdot10^6$ particles' vertical 
velocities (over the entire integration time) is plotted in Fig.~\ref{f:KUR3} 
against the Froude number for the runs in Table \ref{t:runs}.
{It is noteworthy that} the curve $K_{\tilde w}(Fr)$ is characterized by a 
non-monotonic behavior, showing a rapid increase and then decrease of the 
Lagrangian kurtosis in a sharp range between $Fr \approx 0.05$ and $\approx 0.3$, 
with a peak centered around $Fr \approx 0.076$. 
The kurtosis of temperature fluctuations (not shown) also deviates from 3
in the same range of parameters.
This result provides clear evidence that stably stratified turbulent flows are 
characterized by large scale intermittency and strong vertical drafts in a
well defined regime of stratification. 
It is also worth to point out that for all the runs outside the range
$0.05<Fr<0.3$ no large-scale intermittency is detected from Lagrangian 
and Eulerian statistics and the kurtosis of the instantaneous Eulerian PDFs 
does not fluctuate and is constantly close to the Gaussian reference value 
($K_w\approx 3$); this is illustrated in Figures \ref{f:PDF} and \ref{f:KUR} 
for the runs with $Fr=0.03$ and $=0.28$.
One last important remark is that all the runs in the intermittent regime 
identified here have buoyancy Reynolds number $10<R_B<10^2$, whereas 
values of the Lagrangian kurtosis $K_{\tilde w}$ compatible with the Gaussian 
case are found for runs with $R_B \sim 1$ as well as 
$R_B \sim 10^3$ (Table \ref{t:runs}).
%
%
%
To better understand the origin of the strong updrafts and downdrafts in the
vertical velocity, and their variation with $Fr$, we now introduce a simple 
model.
\section{A model for the vertical drafts}
It was found in \cite{rorai_14} that the intermittent behavior observed 
in a stratified fluid can be explained by a 1D model for the vertical velocity 
and the temperature based on {\cite{vieillefosse_84,Li}}. 
In this model, the strong and fast events occur because of an amplification
in the formation of vertical negative velocity gradients resulting from the 
interplay between the time scales of the waves and the nonlinearity. 
We generalize here this model adding both forcing and dissipation, in order to
achieve a steady state and to control the growth of $w$. To this end 
we reduce Eqs.~(\ref{bequationV},\ref{bequationT}) to the 1D case, 
dependent only on the $z$ coordinate, and with ${\bf u}_\perp=0$.  
We derive the equations with respect to $z$ using
 $\partial_z ( \partial_t a + w \partial_z a) = d_t (\partial_z a) + (\partial_z w) \, (\partial_z a)$ 
(for any field $a$, and where $d_t$ is the Lagrangian derivative in the vertical direction), and finally 
we convert spatial derivatives to increments by assuming fields are smooth on an 
arbitrary scale $\ell_z$, and thus $\delta a \approx \partial_z a \, \ell_z$. 
The resulting model has the same overall structure {as} the Boussinesq model, 
Eqs.(\ref{bequationV},\ref{bequationT}), with viscous damping and mechanical
forcing finally re-introduced in an empirical way:
\begin{eqnarray}
\frac{d\delta w}{dt} &=&-\frac{\delta w^2}{\ell_z}-N\delta\theta-\nu\frac{\delta w}{\ell_z^2}+f \\
\frac{d\delta\theta}{dt} &=&-\frac{\delta w\delta\theta}{\ell_z}+N\delta w-\kappa\frac{\delta\theta}{\ell_z^2} \
\label{model2}  \end{eqnarray}
For a fixed value of $\ell_z$ and for high enough $N$, waves prevail over 
turbulence, whereas for small $N$, turbulence prevails over waves. 
At intermediate values, a different behavior occurs, with a rapid increase of 
negative velocity field increments \cite{rorai_14}. This model can be seen as a 
1D approximation to Lagrangian trajectories in a stratified flow, and at fixed 
$N$ with only one free parameter, the length scale $\ell_z$ representing the 
dominant gradients.
The model was integrated using $N$, $\nu$ {and $\kappa$} from the simulations 
(see Table \ref{t:runs}), and for each case we ran an ensemble of 20 realizations (or 
Lagrangian particles) up to $t=20$, using a white-noise random forcing $f$ with 
frequencies between $N/8$ and $N$ to mimic the flat Lagrangian spectrum of the 
vertical velocity observed in stably stratified turbulence, the so-called 
Garrett-Munk spectrum~\cite{ivey_08}. 
{The amplitude of $f$ is the only tunable parameter that is not taken 
directly from the DNS and it was fixed to lead to finite amplitude solutions 
even in the weakly stratified cases (i.e. for moderate $N$)}.
We adjusted the free parameter $\ell_z$ by assuming that the extreme events are 
the result of local shear instabilities or overturning (see Fig.~\ref{f:RENDER}, 
right), and thus the vertical scale at which these events take place can be
expected to be proportional to the Ozmidov scale in each run, 
$\ell_z = \alpha \ell_{Oz}$ (note this scale also separates the boundary between 
wave- and eddy-dominated scales). 
In the following we use $\alpha=4$ which gives the best quantitative agreement 
with the DNS, but any choice for $\alpha$ of order unity gives the same 
qualitative results. Other choices for $\ell_z$, such as $\ell_z = L_B = 2\pi U/N$  
do not give results compatible with the DNS, which can be expected since, at 
the buoyancy scale $L_B$, the vertical Froude number is unity, and thus the 
behavior of the model becomes independent of $Fr$.
Figure~\ref{f:KUR3} (top) shows the kurtosis $K_{\tilde w}$ of the Lagrangian 
vertical velocities for all the DNS runs of this parametric study, 
together with the kurtosis of the Lagrangian field $\delta w$ generated by the 
1D model initialized with the corresponding DNS parameters (red line). 
Each run is characterized by a different Froude number in a range of values 
spanning over one order of magnitude, from $\approx 0.01$ to  $\approx 1$ 
(see Table \ref{t:runs}). Note the behavior of both kurtoses confirms the 
results in Fig.~\ref{f:PDF}: for intermediate values of $Fr$ the vertical 
velocity becomes very intermittent with values of the Lagrangian kurtosis 
$K_{\tilde w}$ larger than $10$.
Considering its simplicity, the model reproduces qualitatively remarkably well 
the behavior found in the DNS. All quantities display a sharp peak for 
$Fr\approx 0.076$, with a  decrease for higher Froude numbers. 
What is striking here is the sharpness of the peak in $Fr$ for all variables, 
as well as the asymmetry between the rising and descending phases of this 
transition.
Such a sharp variation in $Fr\approx 0.076$ may be surprising; it evokes a wave 
resonance  as occurs in critical layers when the flow velocity is equal to 
the wave phase speed, leading to the creation of jets.
Indeed, similarly high values of kurtosis are found in atmospheric data in the 
upper boundary layer, and are interpreted as the signature of a global 
(large-scale) intermittency for a turbulence that is patchy and 
inhomogeneous \cite{mahrt_89}. 
In the model, both the peak and the asymmetry arise as the result of two 
competing effects: for large $Fr$ the nonlinear term dominates, amplifying 
negative velocity increments on a time scale of the turnover time; for small 
$Fr$ the linear term dominates, resulting in wave dynamics. But for intermediate 
values of $Fr$ the time scales of the two terms are similar, with an 
acceleration in the nonlinear amplification resulting from the slightly faster 
wave time scale. To the right and to the left of the peak, the nature of the 
dominant terms is different.
We stress that the transition discussed here strongly differs from 
that observed when the buoyancy Reynolds number, $R_B$, is varied;
this transition affects the 
small-scale flow properties (see, e.g., \cite{shih_05,ivey_08,pouquet_18}). 
\section{Structures, overturning and mixing}
\begin{figure*}  
 \centering    
 \includegraphics[width=18cm]{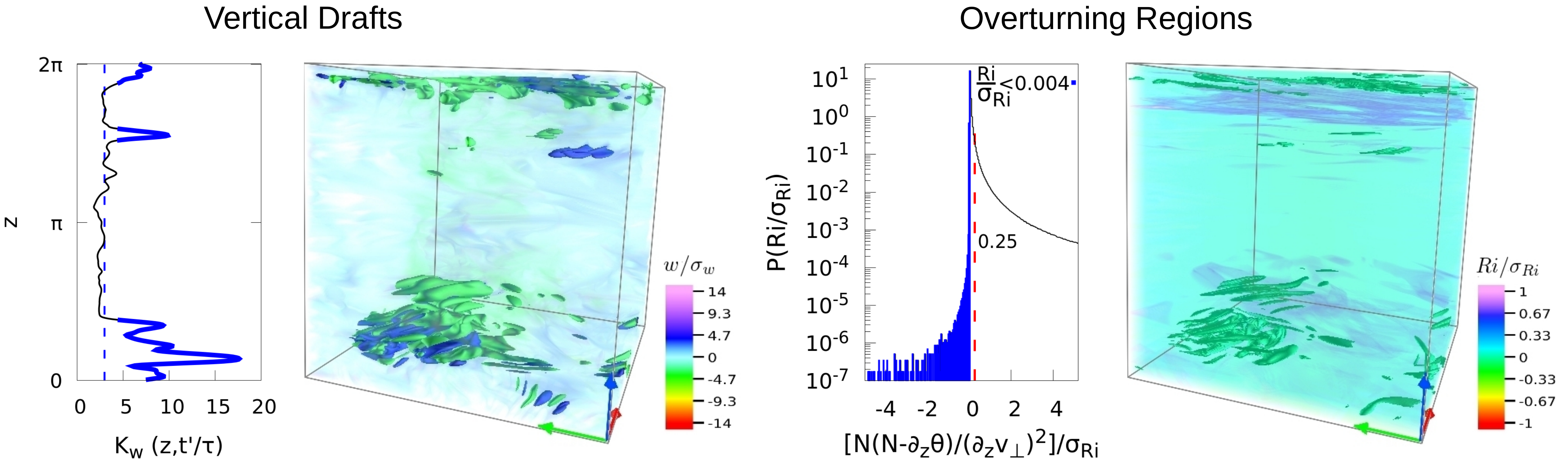}   
 \caption{{\it Left: Variation with height of the kurtosis of the Eulerian vertical 
  velocity $(w)$, computed by plane for run 9 at fixed time $(t'/\tau$, relative 
  maximum of $K_w(t/\tau)$, Fig.\ref{f:KUR}$)$, 
  together with the rendering of $w$ for the same run. A threshold is used to 
  highlight the presence of intense vertical drafts $( > 3 \sigma_w)$
  which appear as large-scale structures emerging in distinct planes that 
  correspond to those with the largest values of the kurtosis. Right: PDF of 
  the gradient Richardson number 
  ({$Ri = N(N-\partial_z\theta)/(\partial_z v_{\perp})^2$}) normalized by 
  its variance $(\sigma_{Ri})$, together with the rendering of its point-wise 
  values where regions prone to develop overturning, with $Ri/\sigma_{Ri}<0.004$, 
  are visualized using opaque colors.}
}
 \label{f:RENDER} 
\vskip-0.1truein      
\end{figure*}

\begin{figure}
 \centering    
 \includegraphics[width=8.6cm]{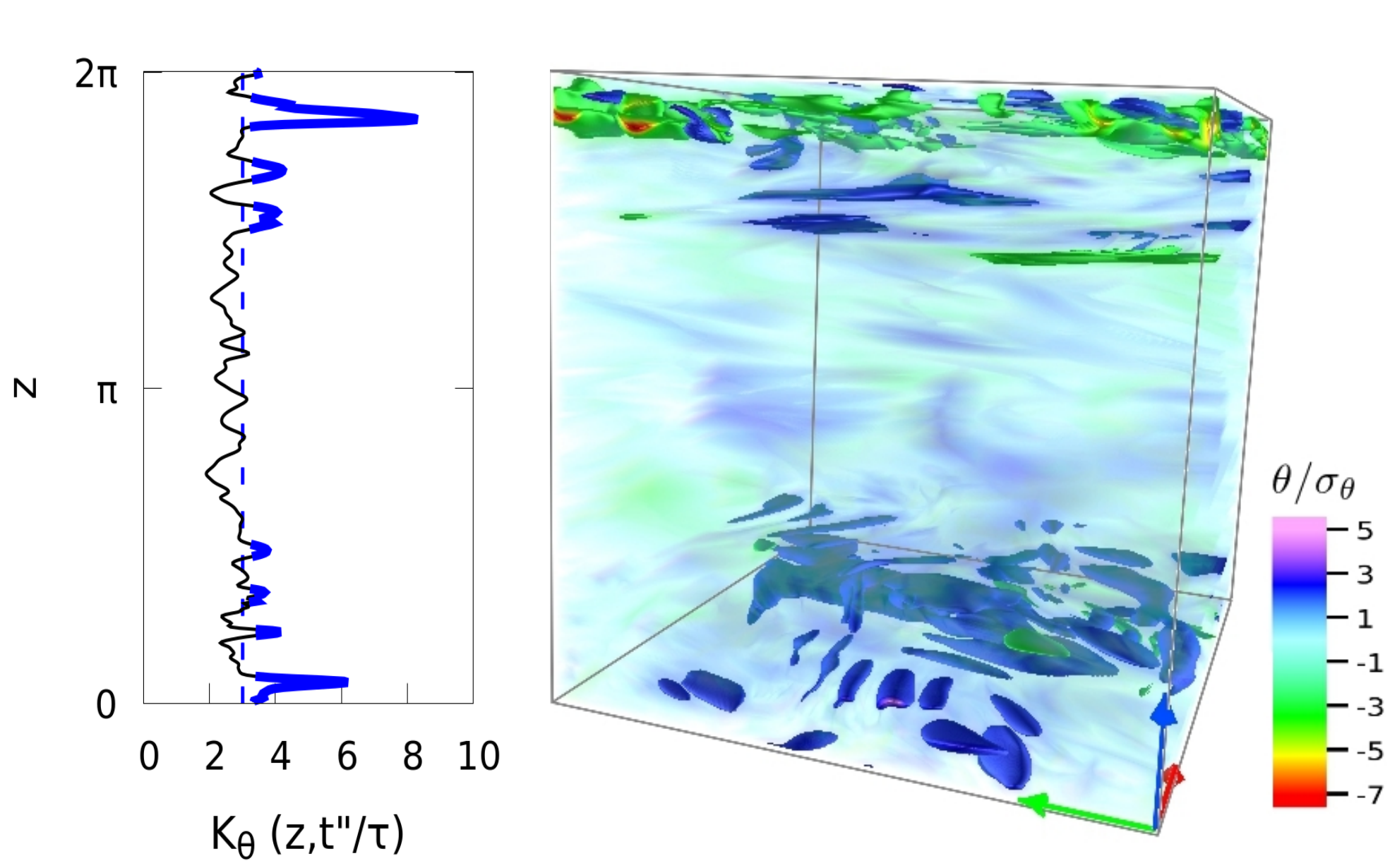}
 \caption{
  {\it Variation with height of the kurtosis of the potential temperature 
  ($\theta$), computed by plane for run 9 at fixed time  $(t''/\tau$, relative 
  maximum of $K_w(t/\tau)$, Fig.\ref{f:KUR}$)$, together with the rendering of 
  $\theta$ for the same run. A threshold is used to highlight with opaque colors
  the presence of hot and cold patches with temperature fluctuations larger than
  3 standard  deviations $(\sigma_{\theta})$, corresponding here to the planes 
  with the largest values of the kurtosis.} 
 }
 \label{f:RENDER2}    
\vskip-0.1truein   
\end{figure}

Thanks to a combined implementation of global spatio-temporal Lagrangian 
statistics and instantaneous Eulerian statistics we are able to show how the
broad distributions observed in previous sections can in fact be linked to 
physical structures.
Fig.~\ref{f:RENDER} (left) provides the three-dimensional rendering of the 
vertical velocity $w$ for run~9, $Fr=0.076$, at the time $t'/\tau$ when a local 
maximum of the instantaneous Eulerian kurtosis is attained based on the curve 
$K_w(t/\tau)$  in Fig.\ref{f:KUR}. 
The side plot in Fig.~\ref{f:RENDER} (left) shows the variation of the kurtosis 
obtained from integrations of the statistics in a single horizontal plane, 
plotted as a function of altitude, $K_w(z,t'/\tau)$, which allow for the 
identification of the planes containing the strongest structures.
On average in the flow, the variable is quasi-Gaussian, with 
$K_w(z,t'/\tau) \sim 3$, except at some rare vertical positions, where the 
fluctuations of $w$ are much larger than the background fluctuations, with 
strong non-Gaussian properties. 
Similar conclusions can be drawn by analyzing the potential temperature $\theta$ 
in physical space, shown in Fig.~\ref{f:RENDER2}. 
This bursty behavior at the large scales, in space and in time, is stronger in 
runs 7 to 12, resulting in the observed non-stationarity of the instantaneous 
Eulerian PDFs (Fig.~\ref{f:PDF}, middle, and Fig.~\ref{f:KUR}) and in the 
emergence of structures, Fig.\ref{f:RENDER} (left) and Fig.~\ref{f:RENDER2}.
The intermittent behavior also affects the smallest scales of the flow: the 
dissipation and the gradient fields also have rare, intense structures embedded 
in the quiet flow (not shown, see \cite{rorai_14}).
%
%
%
%
$ $

Extreme updrafts and downdrafts affect the vertical transport.
The product of the vertical temperature flux with the \BV\ $N$ is the 
so called buoyancy flux $B_f= N\left<w\theta\right>$, routinely 
used to characterize mixing in stratified flows. There are several ways 
to define the mixing efficiency 
(see \cite{venayagamoorthy_16,mashayek_17,pouquet_18} and references 
therein), one possibility being to take the ratio of the buoyancy flux to the 
rate of kinetic energy dissipation $\varepsilon_V$ in the momentum equation. 
Following \cite{maffioli_16l,pouquet_18}, we define here the irreversible mixing 
efficiency $\hat{\Gamma}$ using the potential energy dissipation rate 
$\varepsilon_P = \kappa (|\nabla \theta|^2)$ instead of $B_f$, so one can write 
$\hat{\Gamma} = \varepsilon_P/\varepsilon_V$.
This definition is based on the assumption that the mixing efficiency should 
only account for the irreversible conversion of available potential energy
into background potential energy, quantified by $\varepsilon_P$.

\begin{figure}     
 \begin{center}
   
 \includegraphics[width=8.5cm]{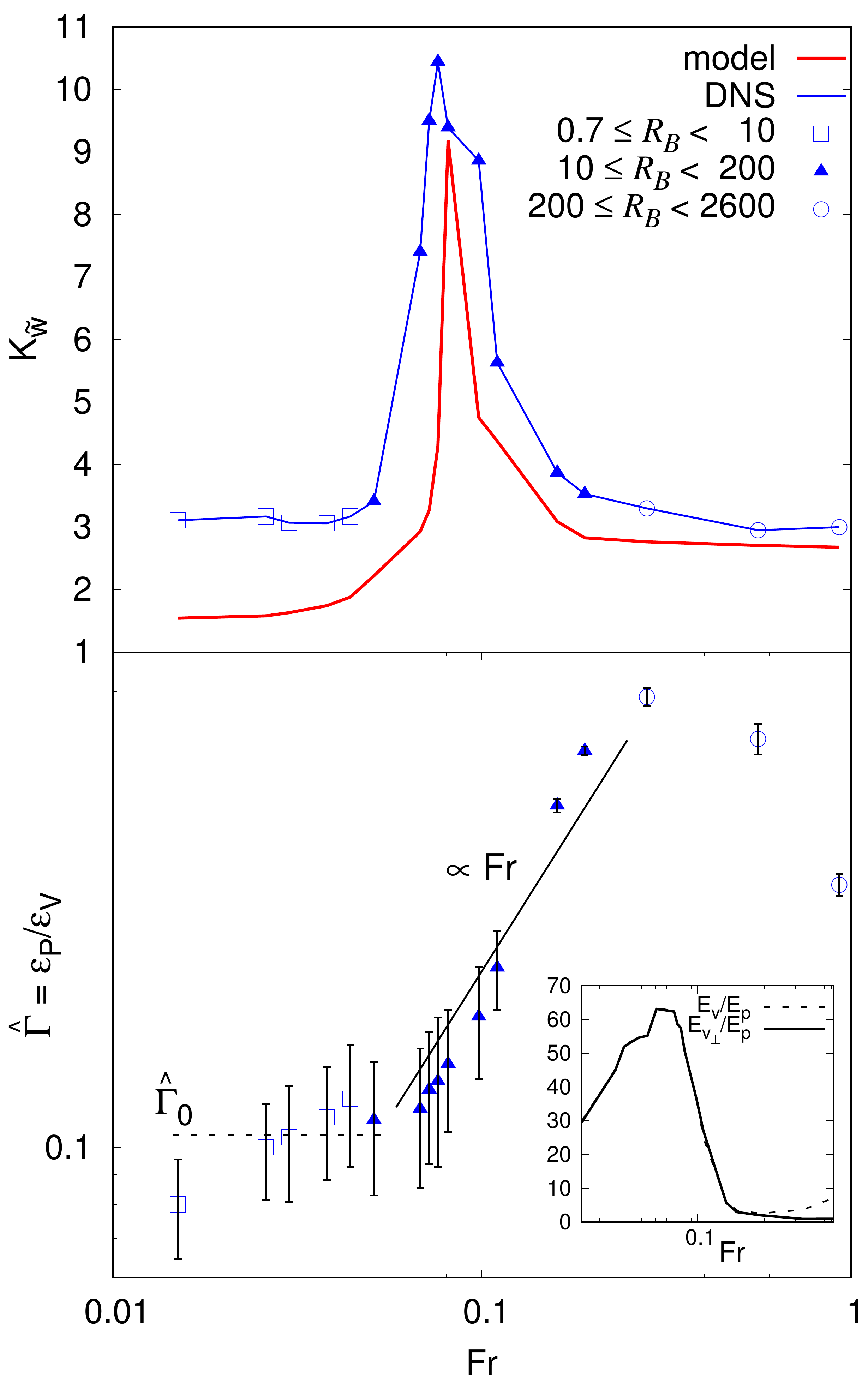}
 \end{center}
\vskip-0.1truein   
 \caption{{\it Top: Kurtosis of the Lagrangian vertical velocity $(\tilde w)$ 
  from the DNS and $(\delta w)$ from the 1D model (red line), as a function of 
  $Fr$.
  Bottom: irreversible mixing efficiency $\hat{\Gamma}$, with 
  $E_v/E_p$ and $E_{v_\perp}/E_p$ in the inset, as a function of $Fr$, all 
  from the DNS; $\hat{\Gamma}_0$ is the estimated saturation value obtained in 
  the limit of small Froude numbers. {The different symbols are used to 
  identify ranges in terms of the buoyancy Reynolds number $R_B$ and highlight
  the non-monotonic trend of $\hat{\Gamma}$ with $Fr$ and $R_B$.}}
}
\label{f:KUR3}   
\vskip-0.1truein   
\end{figure}

$\hat{\Gamma}$ is plotted in Fig.~\ref{f:KUR3} (bottom) against the Froude 
number for all the runs in table \ref{t:runs}. Very interestingly, we find that 
in a parameter space compatible with regions of the atmosphere and the oceans,
namely $0.05<Fr<0.3$, with $R_B > 10$, the irreversible mixing efficiency 
scales linearly with the Froude number ($\hat{\Gamma} \propto Fr$) and it 
increases by roughly one order of magnitude, before dropping for $Fr>0.3$ 
consistently with \cite{maffioli_16l,pouquet_18}.
This is also the range of parameters in which maximal kurtosis of the Lagrangian 
vertical velocity $\tilde w$ and large-scale intermittency attain as a result 
of our study (Fig.~\ref{f:KUR3}, top).
Note that the scaling $\hat{\Gamma} \propto Fr$ observed here for values of 
the Froude number of geophysical interest is different {from the} scaling 
reported in \cite{maffioli_16l} ($\hat{\Gamma} \propto Fr^{-2}$) obtained in the 
case of weak stratification ($Fr>1$). 
For $Fr\ll0.05$ we find a saturation value $\hat{\Gamma}_0\sim 10^{-1}$, 
compatible with the proxy of the mixing efficiency commonly used in the ocean 
community ($\approx 0.2$) \cite{osborn}.
{$\hat{\Gamma}$ exhibits as well a non-monotonic dependence on the buoyancy 
Reynolds number (see table \ref{t:runs}), with a peak value obtained for 
$R_B \sim 200$, consistently with \cite{mashayek_17}}.
To complement our characterization of the mixing, we finally evaluated for all 
runs the ratio of the volume-averaged kinetic to potential energies 
$E_V/E_P$, which can be linked to wave-eddy partition \cite{marino_15}. 
This quantity is plotted in the insert of Fig.~\ref{f:KUR3} (bottom), together 
with the ratio of the volume-averaged horizontal kinetic to potential energies 
$E_{V\perp}/E_P$, versus the Froude number. 
These ratios provide the simplest measure of the partition of energy between 
kinetic and potential modes at all scales, which is another way to estimate the 
efficiency of the mixing. 
Both $E_V/E_P$ and $E_{V\perp}/E_P$ peak in the vicinity of the maximal value 
of $K_{\tilde w}(Fr)$, obtained for $Fr \approx 0.076$ (Fig.~\ref{f:KUR3}, top).
This suggests the possibility that in flows characterized by strong large-scale 
intermittency the mixing enhancement occurs due to large scale overturning, with 
a consequent increase of the kinetic over potential energy (both integrated over 
the volume and in Fourier space). It is also interesting to notice that 
$E_V/E_P \approx E_{V\perp}/E_P$, perhaps due to a more efficient generation 
of horizontal winds in this regime, that makes the horizontal kinetic energy 
dominate the ratio. 

In order to investigate the tendency of stratified flows to develop 
local overturning and the link with the emergence of large scale intermittency 
and structures, we have analyzed the statistics of the point-wise gradient 
Richardson number $ Ri = N(N-\partial_z\theta)/(\partial_z v_{\perp})^2$
computed on the instantaneous Eulerian fields.
This analysis allowed to reveal a clear spatio-temporal correlation between 
the presence of structures in the velocity fields $-$ originating from strong 
vertical drafts $-$ and patches of the flow characterized by values of $Ri$ 
indicating the most unstable regions of the domain. 
As an example, in Fig.~\ref{f:RENDER} we propose the comparison between the 
rendering of the Eulerian vertical velocity $w$ for run~9 $-$ where only 
vertical drafts stronger than three standard deviations $\sigma_w$ are 
visualized with opaque colors (left panel)$-$ and the visualization of the 
point-wise gradient Richardson number for the same run, at the same time 
(right panel). Only the values below the threshold  $0.004$ in the PDF of 
$Ri/\sigma_{Ri}$ (side plot in Fig.~\ref{f:RENDER}, right) are rendered without 
transparency, which makes it very clear the spatial correlation between 
overturning regions and vertical velocity structures in the flow. These 
structures are in turn responsible for the fat wings in the PDFs of the vertical 
Lagrangian and Eulerian velocities we discussed previously. 
This correlation pattern has been observed for runs in the range $0.05<Fr<0.3$ 
and at different times in each run, being more evident in the proximity of the 
local maxima of $K_w(t/\tau)$ (see Fig.~\ref{f:KUR}).  
%
%
%
%
\section{Discussion and Conclusions}
%
We have shown that very strong and intermittent large-scale vertical drafts 
can develop in stratified turbulence in a distinct range of Froude number that 
also reflects an abrupt change in the behavior of the vertical velocity 
statistics and fluid mixing properties. 
Similar results of bursty events in the vertical velocity have been obtained
in \cite{he_15} using DNS of the stable planetary boundary layers with 
comparable $Re$, no-slip boundary conditions in the vertical 
direction of a box with an aspect ratio $\sim 10$.  
These authors find a peak in the 
number of bursting events for Froude numbers 
compatible with the range we identified here $(0.05<Fr<0.3)$ in which we observe 
large-scale intermittency with a peak of the kurtosis of the Lagrangian vertical 
velocity ($K_{\tilde w}$) obtained for $Fr \approx 0.76$. Bursts in stably 
stratified turbulence were also reported in \cite{rorai_14}. 
In the former case, the observed intermittency is mostly associated with
interactions with the boundary, a situation which we do not have in the present 
study, while in the latter, albeit for only two values of $Fr$, similar 
extreme events were found. Also, in analysis of climatological data in the free 
troposphere, it was found that the fields with the strongest departure from 
Gaussianity are the vertical velocity, together with the specific humidity, 
and it has been speculated that extreme events in climate behavior such as recent 
heat waves may be linked to specific resonances in Rossby wave dynamics 
\cite{petoukhov_16}.
These results, however, lead to difficulties when mixing efficiency is 
parameterized. As reported in \cite{gregg_18}, results from experiments and 
numerical simulations attempting characterization of the mixing efficiency 
fail to converge in many cases. 
This may be related to the strong non-stationarity of the data for
intermediate and small values of $Fr$, as we have shown in Fig.\ref{f:KUR} 
($Fr=0.076$) for the initial phase of evolution of the flow, well beyond 
the peak of the dissipation.

The present work provides evidence that even without the added complexity 
of climate processes (including moisture, boundaries and topography) 
extreme events can be observed in both the vertical velocity and the
temperature fluctuations. We also showed how these events are linked to the 
emergence of structures which in turn correlate, in time and space, with 
those regions where the flow is more unstable and prone to develop overturning. 
Our results show that this behavior takes place in a range of Froude numbers 
relevant for atmospheric and oceanic flows and for values of the buoyancy Reynolds 
number $> 10$.
{Even though the values of $Re$ considered remain small compared to the 
troposphere and the ocean, they are not too far off for the mesosphere lower 
thermosphere (MLT) \cite{chau, marino_15}. Recent numerical studies point also 
to the possibility that above certain thresholds in $R_B$, trends for the mixing 
efficiency and dissipation represent realistic scenarios even if obtained at $Re$ 
smaller than those observed in nature \cite{pouquet_18,marino_15a}}. 
Another important result of this Letter is that the observed behavior can be 
reproduced with a simple 1D model which is a truncation of the full system of 
governing equations in the Boussinesq framework; this indicates that the 
the large scale intermittency range in $Fr$ corresponds to a region in the
parameter space in which the time scales of waves and nonlinearities near the 
Ozmidov length-scale are comparable, thus resulting in fast resonant 
amplification of velocity differences. 
Finally we found that the irreversible mixing efficiency parameter increases by 
roughly one order of magnitude (from $\hat{\Gamma} \sim 0.1$ to $\hat{\Gamma} \sim 1$) and 
scales linearly with the Froude number ($\hat{\Gamma} \propto Fr$) in the range 
$0.05<Fr<0.3$, relevant for geophysical flows \cite{gregg_18}. 
%
%
\end{document}